\documentclass[sigconf]{acmart}

\usepackage{xspace}
\usepackage{textcomp}
\usepackage{soul}
\usepackage{marginnote}
\usepackage{hyperref}
\usepackage{amsfonts}
\usepackage{graphicx}
\usepackage{float}
\usepackage{subfig,rotate}
\usepackage{verbatim}
\usepackage{comment}
\usepackage{versions}
\usepackage{multirow}
\usepackage{longtable}
\usepackage{booktabs}
\usepackage{makecell}
\usepackage{caption}
\usepackage{acro}
\usepackage{tabularx}
\usepackage{adjustbox}
\usepackage{listings}
\usepackage{color}

\def\bitcoin{%
  \leavevmode
  \vtop{\offinterlineskip 
    \setbox0=\hbox{B}%
    \setbox2=\hbox to\wd0{\hfil\hskip-.03em
    \vrule height .3ex width .15ex\hskip .08em
    \vrule height .3ex width .15ex\hfil}
    \vbox{\copy2\box0}\box2}}

\definecolor{dkgreen}{rgb}{0,0.6,0}
\definecolor{gray}{rgb}{0.5,0.5,0.5}
\definecolor{mauve}{rgb}{0.58,0,0.82}

\acmConference[Technical Report]{}{July}{2019}
\setcopyright{none}
\renewcommand\footnotetextcopyrightpermission[1]{}

\begin{document}
\title[Deanonymizing Tor hidden service users through Bitcoin transactions analysis]{Deanonymizing Tor hidden service users through Bitcoin transactions analysis}

\author{Husam Al Jawaheri}
\affiliation{
  \institution{Qatar University}
}
\author{Mashael Al Sabah}
\affiliation{
  \institution{Qatar Computing Research Institute, HBKU}
}
\author{Yazan Boshmaf}
\affiliation{
  \institution{Qatar Computing Research Institute, HBKU}
}
\author{Aiman Erbad}
\affiliation{
  \institution{Qatar University}
}
\renewcommand{\shortauthors}{Al Jawaheri et al.}

\begin{abstract}
With the rapid increase of threats on the Internet, people are continuously seeking privacy and anonymity. Services such as Bitcoin and Tor were introduced to provide anonymity for online transactions and Web browsing. Due to its pseudonymity model, Bitcoin lacks retroactive operational security, which means historical pieces of information could be used to identify a certain user. We investigate the feasibility of deanonymizing users of Tor hidden services who rely on Bitcoin as a payment method by exploiting public information leaked from online social networks, the Blockchain, and onion websites. This, for example, allows an adversary to link a user with @alice Twitter address to a Tor hidden service with private.onion address by finding at least one past transaction in the Blockchain that involves their publicly declared Bitcoin addresses.

To demonstrate the feasibility of this deanonymization attack, we carried out a real-world experiment simulating a passive, limited adversary. We crawled 1.5K hidden services and collected 88 unique Bitcoin addresses. We then crawled 5B tweets and 1M BitcoinTalk forum pages and collected 4.2K and 41K unique Bitcoin addresses, respectively. Each user address was associated with an online identity along with its public profile information. By analyzing the transactions in the Blockchain, we were able to link 125 unique users to 20 Tor hidden services, including sensitive ones, such as The Pirate Bay and Silk Road. We also analyzed two case studies in detail to demonstrate the implications of the resulting information leakage on user anonymity. In particular, we confirm that Bitcoin addresses should always be considered exploitable, as they can be used to deanonymize users retroactively. This is especially important for Tor hidden service users who actively seek and expect privacy and anonymity.
\end{abstract}

\maketitle

\section{Introduction}
\label{sec:introduction}
Anonymity and privacy over the Internet are becoming more critical than ever. For that, many solutions are being deployed to improve the anonymity of users while making online transaction or browsing the web. The most famous of these solutions are the decentralized cryptocurrencies and Tor anonymity network. One of the early examples is the Bitcoin network~\cite{nakamoto:2008}, which provides users with the ability to perform online transactions "pseudonymously". Due to its popularity, more than 100K merchants worldwide accept Bitcoin payments~\cite{cuthbertson_2015}. One of the reasons of Bitcoin's popularity is its presumed anonymity. Tor~\cite{tor} is the most widely used anonymous communication network with millions of daily active users~\cite{tor_metric}. In addition to client-side privacy and anonymity, Tor also enables server-side anonymity through the design of hidden services. The goal of hidden services is to safely enable online freedom, anticensorship, and end-to-end anonymity and security~\cite{hs_gooduse}. Indeed, for
those reasons, hidden services are being operated by whistleblowing websites such as WikiLeaks,\footnote{\url{https://wikileaks.com}} search engines such as DuckDuckGo,\footnote{\url{https://duckduckgo.com}} and online social networks such as Facebook.\footnote{\url{https://facebook.com}} Hidden services have also become a breeding ground for Dark Web vendors, such as Silk Road~\cite{christin2013traveling} and Agora~\cite{van2016sells}, which offer illicit merchandise and services~\cite{HScontent,Moore}.

As discussed by Vincent and Johan~\cite{MieghemP15}, Tor and Bitcoin represent the main components needed to achieve anonymous online purchases with exhaustive operational security. In this context, operational security is the process of protecting individual pieces of information that could be used to identify a user. Unfortunately, Bitcoin lacks retroactive operational security due to its pseudonymity model~\cite{nakamoto:2008}. This model has an important limitation because of the linkability of Bitcoin transactions that are stored in the Blockchain and their public availability.

\paragraph{Problem.}
Due to potential information leakage, using Bitcoin as a payment method is a serious threat to the anonymity of Tor hidden services and their users. Yet, Bitcoin is the most popular choice for these services for accepting donations or selling merchandise~\cite{HScontent}. Moore and Rid~\cite{Moore} studied how hidden services are used in practice, and noted that Bitcoin was the dominant choice for accepting payments for these services. Although multiple studies~\cite{fleder:2015, meiklejohn:2013, DuPont:2015} demonstrated that Bitcoin transactions are not as anonymous as previously thought, Bitcoin remains the most popular digital currency on the Dark Web~\cite{castillo_2016}, and many users still choose to use it despite its false sense of anonymity. Biryukov et al.~\cite{tor_bitcoin-nono} showed that even if users use Bitcoin over an anonymity network such as Tor, they are still vulnerable to deanonymization and man-in-the-middle attacks at the network level. While previous studies analyze the vulnerabilities that result from using Bitcoin over Tor~\cite{tor_bitcoin-nono}, mostly at the network level, we provide the first study that focuses on the application level, shedding light on an exploitable information leakage resulting from correlating public profiles of online social network users with Bitcoin transactions and onion websites.

Hidden service users are one class of Bitcoin users whose anonymity is particularly important. Hidden service operators and users are actively seeking to maintain their anonymity. However, those users are under the risk of deanonymization when they reveal their Bitcoin addresses. By studying the transactions associated with these addresses, a significant amount of information can be leaked and used to collect sensitive information about hidden services and their customers, where a user can be linked to a hidden service.

In this paper, we seek to understand the privacy risk associated with using Bitcoin as a payment method by Tor hidden services. That is, we show how using Bitcoin leaks public information that can be exploited to deanonymize Tor hidden service users. In particular, we demonstrate the feasibility of linking a user with @alice social network profile to a Tor hidden service with private.onion address. We note that this research has been carefully revised and approved by our institution's IRB board, and it does not put users at any risk other than what they are currently exposed to (Section~\ref{sec:ethics}). We also discuss a number of countermeasures that are designed to protect and improve user privacy and anonymity (Section~\ref{sec:countermeasures}).

\paragraph{Approach.}
By browsing onion landing pages of various hidden services, we observed that it is possible to extract their payment Bitcoin addresses from static HTML content. Accordingly, we crawled 1.5K hidden service pages and created a dataset of 88 Bitcoin addresses operated by those hidden services, including two ransomware addresses. We also crawled online social networks for public Bitcoin addresses, namely, Twitter\footnote{\url{https://twitter.com}} and the BitcoinTalk forum.\footnote{\url{https://bitcointalk.org}} Out of 5B tweets and 1M forum pages, we created two datasets of 4.1K and 41K Bitcoin addresses, respectively. Each address in these user datasets is associated with an online identity and its corresponding public profile information (Section~\ref{subsec:data_collection}).

Using a clustering heuristic proposed by Meiklejohn et al.~\cite{meiklejohn:2013}, we performed a wallet-closure analysis that allowed us to expand the collected Bitcoin addresses per user. In other words, for each address in the collected user datasets, we identified other addresses belonging to the same user who owns the address. This closure analysis approximates a user's wallet, which is the set of addresses that are controlled by the user, and thereby increase the number of identified links between users and hidden services. One problem with closure analysis is that the closure can over-approximate the size of the wallet, as a consequence of mixing~\cite{mixingservices} and CoinJoin~\cite{coinjoin} services. Therefore, we excluded closures that have common addresses from the analysis. This ensures that users are not double-counted and reported results are lower-bound estimate, as each remaining closure represents a subset of a wallet whose addresses are controlled by a unique user (Section~\ref{subsec:closure_analysis}).

After wallet-closure analysis, we used the datasets to analyze the transactions in the Blockchain. In particular, we searched for transactions between user and hidden service addresses in order to identify links between them. This enabled us to associate users, or online identities, with hidden services and access their transaction history. To demonstrate the impact of linking, we traced and described two case studies showing that Bitcoin addresses can be used to deanonymize users retroactively, starting from May, 2010. It is important to highlight that deanonymization depends solely on information leaked from public data sources. Finally, to gain insights into the economic activity of the linked hidden services, we analyzed the corresponding transaction history, focusing on number of transactions, the amount of money being exchanged, and the lifetime of these hidden services (Section~\ref{subsec:api}).

\paragraph{Results.}
With wallet-closure analysis, we were able to expand the datasets from 45.2K Bitcoin addresses to more than 19.1M, with an average of 425 addresses per user. Using transaction analysis, we were able to link 125 unique users to 20 Tor sensitive hidden services, such as WikiLeaks, Silk Road, and The Pirate Bay.\footnote{\url{https://thepiratebay.org}} The case studies unmasked multiple users of The Pirate Bay hidden service, along with their personally identifiable information (PII), such as name, gender, age, and location.\footnote{In accordance with our IRB board's guidelines, we have removed the PII of these users, as the linked hidden services are considered illegal in their countries.} We also found that users from multiple countries and different ages had links with the Silk Road address in our hidden service dataset. One of the users, for example, is a teenager who has many social network accounts showing his real identity.

The economic activity analysis of the linked hidden services shows that WikiLeaks is the highest receiver of payments in terms of volume, with 25.6K transactions. In terms of the amount of incoming payments, however, the Silk Road tops the list with more than 29.6K Bitcoins received on its address.\footnote{As of Dec 2017, this amount of Bitcoins equals more than 580 million USD.} We observed that the money flowing in and out of hidden service addresses is nearly the same. This suggests that these services do not keep their Bitcoins on the address they use for receiving payments, but rather distribute the coins to other addresses instead. Finally, from the last transaction dates on the addresses, we found that only eight out of the 20 linked hidden services are inactive in 2017. This, however, does not mean the inactive services stopped making Bitcoin transactions, as they might have used different addresses that we do not know.

\paragraph{Contributions.}
This paper shows the implications of Bitcoin's pseudonymity model, which lacks retroactive operational security, on Tor hidden service users. Our contributions are the following:
\begin{enumerate}
\item A method that links online user identities with Tor hidden services through Bitcoin transactions analysis. The method improves linking results by
using closure analysis techniques and by significantly eliminating the noise from mixing and CoinJoin services.
\item The first real-world experiment that shows the feasibility of deanonymizing Tor hidden service users by exploiting information leakage resulting from correlating public data sources, namely, online social networks, the Bitcoin Blockchain, and Tor hidden services.
\item An economic activity analysis of 20 hidden services that were used by linked users. This includes statistics on their transaction volume, flow of money, and lifetime.
\end{enumerate}

\section{Background}
\label{sec:background}
We now present the necessary background on Bitcoin and Tor.

\subsection{Bitcoin}
Bitcoin~\cite{nakamoto:2008} is a decentralized digital crypto-currency system which eliminates the need for a central bank authority to manage the transfer of funds. The Bitcoin network is maintained by a peer-to-peer network of miners who validate transactions without relaying on trust. Due to its popularity, more than 100K merchants worldwide accept Bitcoin payments~\cite{cuthbertson_2015}. One of the reasons of Bitcoin's popularity is its presumed anonymity.
The identities of users in Bitcoin are hidden using pseudonyms which are used as addresses to perform transactions. A Bitcoin address is a 160-bit hash of a public key generated by a digital signature algorithm. The set of public/private keys that are owned by a user is called a wallet. Private keys are used to sign inputs of transactions as a proof of ownership.

\subsubsection{Transactions.}
Alice makes a payment to Bob by creating a new transaction. She uses one or more Bitcoin addresses that she controls as inputs. She also includes the amount to be transferred, and chooses Bob's address(es) as a transaction output. To protect the transaction, she signs it using her private key(s), and then broadcasts it to the whole network. In order to verify transactions, and be rewarded with new generated coins, miners collect the broadcast transactions, embed them in a well-defined data structure called a block, and then attempt to solve a hashing computational puzzle involving the block. When the block is solved, it is attached to the Blockchain, which is a hash-chain that maintains all solved blocks, and thereby all embedded transactions ever created and verified in the Bitcoin network.

The Blockchain is publicly maintained and can be downloaded over the Internet, Bitcoin's original client, or explored using centralized servers, such as Blockchain Info.\footnote{\url{https://blockchain.info}} Every transaction in the Blockchain has a list of inputs and outputs, where each includes addresses that were used in the transaction and the amount of coins spent in that transaction. Transactions downloaded from Blockchain Info include more information, such as the relay IP address and the transaction timestamp that records the time at which the transaction was made.

\subsubsection{Anonymity.}
While transactions in Bitcoin are presumed to be anonymous, linkability between addresses is possible due the nature of the Blockchain~\cite{nakamoto:2008}. For example, one can verify if Alice and Bob have a transaction between them. Furthermore, if Alice owns multiple addresses, one may be able to link them as belonging to the same person.

Meiklejohn et al.~\cite{meiklejohn:2013} observed that two Bitcoin addresses, $A$ and $B$, belong to the same user if both $A$ and $B$ have been used as inputs for the same transaction, or $B$ receives, as an output, the unspent change of a
transaction where $A$ is used as input. The authors used this observation to define a heuristic for mapping multiple addresses to an entity representing a unique user. Specifically, the heuristic is based on the idea that since the private keys of the user are used to sign the inputs $A$ and $B$, then both $A$ and $B$ are controlled by the same person. As the addresses or the underlying public/private keys that are owned by a user represents a wallet, the heuristic tries to induce the wallet of a user given a subset of the addresses in the wallet. The authors also define a second heuristic based on another observation. When an address is used as an input in a transaction, all of its associated Bitcoins have to be spent at once. If those coins exceed what the sender wants to spend, then the sender has to reference two outputs, one to the receiver with the intended amount, and another for the change. The sender typically controls the change address within the transaction. Both heuristics represent wallet-closure techniques that are used for Bitcoin transaction analysis.

It is important to note that wallet-closure techniques are noisy and can result in addresses that do not belong to the same user or wallet. One reason for this is the use of mixing~\cite{mixingservices} and CoinJoin~\cite{coinjoin} services. Given a set of input addresses of multiple users, these services generate a sequence of transactions that effectively mixes the coins to enhance anonymity. In this work, we modify the first wallet-closure technique to handle Bitcoin mixing for transaction analysis, as described in Section~\ref{subsec:closure_analysis}.

\subsection{Tor Hidden Services}
Tor~\cite{tor} is the most widely used anonymous communication network available online. Tor enables server-side anonymity through the design of hidden services, also known as onion services. To achieve their anonymity goal, a hidden service client and operator establish a communication tunnel, known as a circuit, between each other over multiple intermediate routers. Anonymity is maintained as long as the intermediate routers at the two ends of the tunnel are not controlled by an adversary who can use time or traffic analysis to link the source to the destination. Hidden services have also been subjected to active attacks in the wild~\cite{hs_wild1,hs_wild2}. For these reasons, the Tor project is actively working on addressing the security weaknesses of hidden services~\cite{hackfest}.

To ensure transaction anonymity, Bitcoin has become the most popular choice by Tor hidden services for accepting donations or selling merchandise~\cite{HScontent}. Unfortunately, this has contributed to the rise of illegal hidden services, such as Silk Road and Agora, which offer illicit merchandises and services~\cite{christin2013traveling,van2016sells,HScontent,Moore}.

\section{Approach and Experiment}
\label{sec:approach}
While the goal of using Bitcoin for Tor hidden services is to provide transaction and browsing anonymity, we show that this usage typically leaks information that can be used to deanonymize hidden service users. In particular, the adversary can link users, who publicly share their Bitcoin addresses on online social networks, with hidden services, which publicly share their Bitcoin addresses on onion landing pages. This is achieved by inspecting historical transactions involving these two addresses in the Blockchain. In doing so, the adversary only relies on data that is publicly available.

\subsection{Overview}
\label{sec:overview}

To illustrate the deanonymization attack, let us consider Alice, a privacy-savvy user who uses Tor, in the following scenario:
\begin{enumerate}
\item Alice uses a browser and creates an online identity @alice with a public profile on social network public.com.
\item Alice uses @alice to make a public post asking for donations on Bitcoin address $A$.
\item Alice receives donations through a number of Bitcoin transactions, where $A$ is used as an output address.
\item Alice uses Tor browser to visit hidden service private.onion that has public Bitcoin address $P$.
\item Alice makes a payment $A\rightarrow P$ to private.onion using $A$ as an input address and $P$ as an output address.
\end{enumerate}

While steps 1--3 involve non-anonymous web browsing and public activities, Alice has some expectations of privacy and anonymity in steps 4--5, given that she is using Tor and Bitcoin. Step 5, however, leaks a piece of information, the transaction $A\rightarrow P$ in particular, that can be used by an adversary, Trudy, to link @alice to private.onion, as follows:
\begin{enumerate}
\item Trudy crawls public.com on regular basis, storing public user profiles and posts.
\item Trudy crawls hidden services on regular basis, storing accessible onion pages.
\item Trudy parses crawled data on regular basis, searching for Bitcoin addresses.
\item Trudy parses the blockchain on regular basis, searching for transactions between user and hidden service addresses.
\item Trudy finds Bitcoin address $A$ on public.com, associated with online identity @alice.
\item Trudy finds Bitcoin address $P$ on private.onion.
\item Trudy finds transaction $A\rightarrow P$ and accordingly links @alice to private.onion.
\end{enumerate}

Unknown to Alice, Trudy can effectively deanonymize Alice's identity and link her to activities on private.onion using steps 1--7. More importantly, this attack vector is feasible retroactively in the future, starting from the time of the transaction $A\rightarrow P$.

\subsection{Adversary Model}
\label{sec:adversary_model}
We assume a passive, limited adversary. The adversary has access to or is capable of collecting Bitcoin addresses of Tor hidden services and their users. This attacker does not need to control network resources, but can extract publicly accessible information from online social networks, the Blockchain, and onion pages. Obtaining Bitcoin addresses of users can be either targeted or non-targeted, depending on the attack scenario. For the earlier scenario, the adversary can use social engineering or exploit contextual metadata. For example, if Trudy knows that Alice booked a ticket on Expedia\footnote{\url{https://www.expedia.com}} at a certain time with a certain amount of coins, Trudy can easily deduce Alice's Bitcoin address from the Blockchain. For the latter scenario, the adversary can crawl and parse public data sources for Bitcoin addresses and associated identities on a large scale.

We focus on the second, non-targeted attack scenario and show that an adversary can deanonymize hidden service users by correlating public data from online social networks, the Blockchain, and Tor hidden services.

\begin{table*}
\centering
{
\begin{tabular}{llrrr}\toprule
& Date collected & & \multicolumn{2}{c}{\# addresses}\\
\cmidrule(rr){4-5}
Label & (dd/mm/yyyy) & \# users &  Original & Expanded\\ \midrule
{\sffamily hiddenServices} & 27/01/2016 & 88 & 88 & --\\
{\sffamily twitterUsers} & 30/12/2014 & 4,183 & 4,183 & 623,189\\
{\sffamily forumUsers} & 26/10/2016 & 40,970 & 40,970 & 19,213,141\\
\bottomrule
\end{tabular}
}
\caption{Dataset summary of collected Bitcoin addresses}
\label{table:datasets}
\end{table*}

\subsection{Ethical Considerations}
\label{sec:ethics}
The deanonymization presented in this work depends on correlating public Bitcoin addresses of users with the transactions revealed by the Blockchain. Many prior studies performed similar analyses based on crawled public Bitcoin addresses~\cite{Reid:2013, fleder:2015, meiklejohn:2013}. While our study narrows down this analysis to the scope of hidden services and their users, we stress that even the Bitcoin addresses of hidden services were readily available on their onion landing pages. We did not try to obtain Bitcoin addresses of hidden services that require authentication, payment, or exchange of emails.

We believe the data collected and used herein is easily available to adversaries. In this research, in addition to the Blockchain and onion landing pages, we used data available from two online social networks, namely, Twitter and the BitcoinTalk forum. A web search engine, such as Google,\footnote{\url{https://google.com}} or any other organization that has access to a significantly larger amount of data could perform the analysis on a larger scale, and potentially exploit a significantly larger amount of leaked information about users. Ignoring the existence of the data, or the security implications of using Bitcoin as a payment method for hidden services, can leave both the users and the security community unaware of the involved privacy leaks.

To this end, we have consulted and received the approval of our institution's IRB board to conduct our experiment. We would like to highlight that our research does not put users at any additional risk, but rather expose the existing one. This is important because once users become vulnerable to this deanonymization attack, as described in Section~\ref{sec:overview}, they stay vulnerable even after they switch to a more secure payment method or stop using the service. As part of our personal and institutional code of ethics, we have reached out to vulnerable users in our datasets and informed them about this threat and possible remedies. We also posted an anonymous notice on BitcoinTalk forum.\footnote{\url{https://bitcointalk.org/index.php?topic=2602885}} In Section~\ref{sec:countermeasures}, we discuss a number of countermeasures that are designed to protect and improve user privacy and anonymity.

\subsection{Data Collection}
\label{subsec:data_collection}
We now describe how we collected public Bitcoin addresses and online identities of Tor hidden services and Bitcoin users.\footnote{The resulting datasets are available per request to interested researchers.}

\subsubsection{Hidden Services.}
\label{sec:data_hidden_services}
Tor hidden services are not indexed by normal search engines, but can be found using indexing services such as Ahmia,\footnote{\url{https://ahmia.fi}} which is accessible from the normal Web. Other search engines are available but require a Tor browser in order to access them. These search engines are used to access the onion landing pages, or the websites, of many hidden services. Typically, hidden services publish their Bitcoin addresses on their landing pages for receiving payments. These addresses can be collected by simply downloading these pages and searching for Bitcoin addresses using regular expressions. As a Bitcoin address is a base-58 encoded identifier of 26--35 alphanumeric characters, beginning with the number 1 or 3, we used the following regex:
\begin{center}
\texttt{*[13][a-km-zA-HJ-NP-Z1-9]\{25,34\}}
\end{center}

With the goal of collecting long-term hidden service addresses, we started expanding the dataset in mid 2015. Over time, however, we found that fewer hidden services publicly exposed their Bitcoin addresses on their onion pages, resorting to online wallets or other crypto-currencies, possibly due to the increasing awareness of Bitcoin's privacy issues  (Section~\ref{sec:discussion}). Therefore, our analysis focuses on the time window when publishing long-term Bitcoin addresses was a common practice. We note that the deanonymization attack is feasible using historic data that is publicly available since the inception of Bitcoin in 2010 up until now.

In our experiment, we first compiled a list of onion addresses from Ahmia. We then downloaded the landing pages of more than 1.5K hidden services. While our goal was to automate the process of collecting Bitcoin addresses, many of the onion addresses listed by Ahmia were unavailable or offline when we ran the scripts on Jan 27, 2016. A simple regex search on the landing pages allowed us to extract a small number of Bitcoin addresses, less than 20 addresses.

Furthermore, by browsing various hidden services, we were able to extract more addresses. We also observed that many services did not expose their Bitcoin addresses on their landing pages, and would require users to attempt purchasing items before a Bitcoin address is shown to the user.\footnote{Services we manually visited offered variety of different content ranging from dark markets (e.g. drug, stolen card, and arms) to services such as WikiLeaks.} In addition, we included two known ransomware addresses that are published on the Web and the Blockchain.\footnote{https://blockchain.info/address/1AEoiHY23fbBn8QiJ5y6oAjrhRY1Fb85uc} Ransomware is a malware category that limits the access of users to their files by encrypting them~\cite{trend_micro}. Ransomware requires victims to pay in order to get access to the decryption keys. To remain anonymous, ransomware requires victims to pay through the Bitcoin network. Ransomware lockers are known to use Tor hidden services as a place to hide their malicious activities~\cite{kimayong_2016}.

Our automated and manual searching resulted in a total of 105 Bitcoin addresses. We verified that those addresses were active by downloading their transactions. We removed all addresses that contained no transactions or had very low amount of Bitcoins, less than 0.00001\bitcoin, and are likely to be inactive. This resulted in 88 unique Bitcoin addresses which represent the {\sffamily hiddenServices} dataset, as summarized in Table~\ref{table:datasets}.

While the number we ended up with seem relatively small compared to the total number of hidden services, our goal here is to show the feasibility of linking hidden services to their users using only public information. As described in Section~\ref{sec:adversary_model}, an adversary that has a wider access to resources, or actively interact with hidden services, is expected to collect a significantly larger number of addresses.

\subsubsection{Users.}
\label{subsubsec:data_collection_users}

Bitcoin users often post their addresses on online social networks for different purposes, such as receiving donations, offering services, or showing that they are part of the community. Public Bitcoin addresses exposed online could potentially put these users at the risk of transactions history tracing and linkage. Not only do users reveal their public Bitcoin addresses, but they also reveal personal information representing their online identities, such as contact information, gender, age, and location, depending on the social network used.

\begin{figure*}
\centering
    \subfloat[{\sffamily twitterUsers}]{\includegraphics[width=0.3\linewidth]{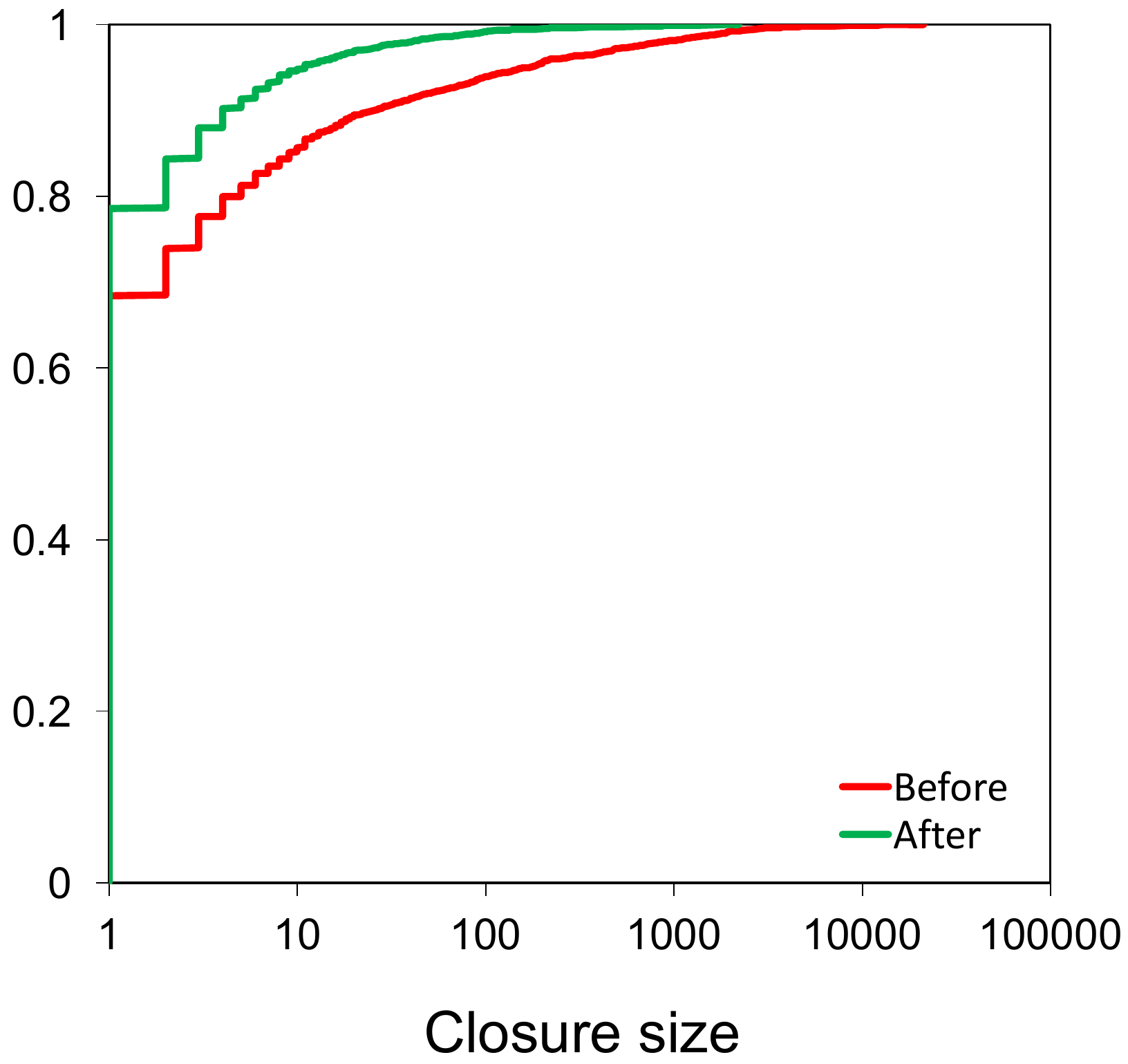}}
    \hspace{1.65cm}
    \subfloat[{\sffamily forumUsers}]{\includegraphics[width=0.3\linewidth]{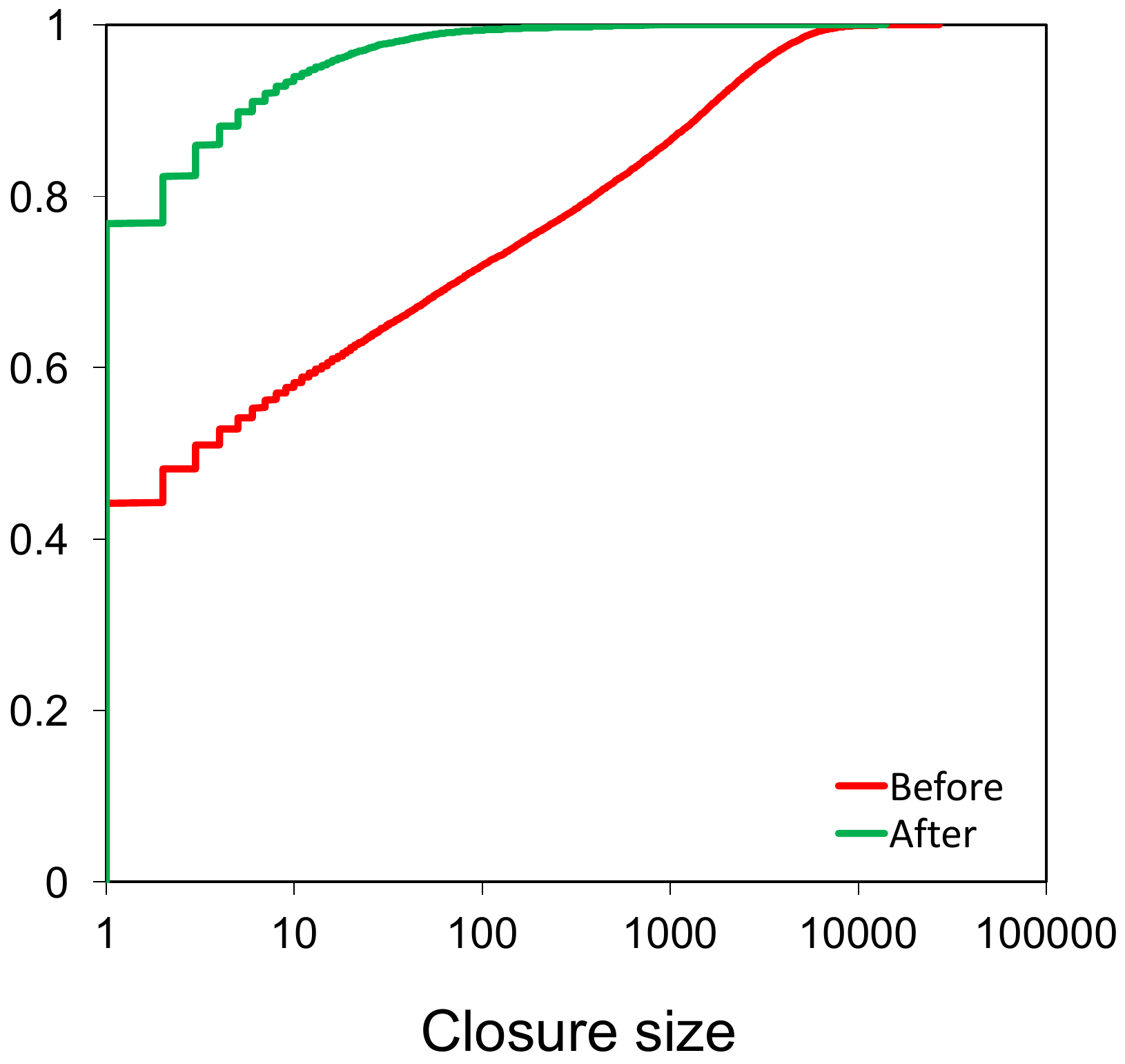}}
\caption{CDFs of closure size before and after cleaning}
\label{fig:closure_cdf}
\end{figure*}

Bitcoin addresses and the associated online identities of users can be collected by crawling and parsing their user profiles or by using the native APIs provided by the social network itself. In our experiment, we collected the addresses and identities of users of two online social networks, specifically, Twitter and the BitcoinTalk forum, as summarized in Table~\ref{table:datasets}. We note that there is a one-one mapping between an address and its associated online identity.

\paragraph{Twitter}
We used Twitter Decahose stream data~\cite{decahose} that we previously collected from Dec 11, 2013 to Dec 30, 2014. Decahose provides a 10\% realtime random sampling of all public tweets through a streaming connection. The reason we chose this dataset is because we wanted to find Bitcoin addresses for Twitter users, which coincide with our hidden services' Bitcoin addresses. Recall that prior to 2016, it was more common for users and hidden services to share their long-term addresses. Overall, data collection resulted in 10TB of JSON-formatted files representing 5 billion tweets. In addition to its textual content, each tweet has the public profile information of its author, which sometimes contains the author's Bitcoin address. To extract tweets that contain Bitcoin addresses, we scanned the whole dataset and kept the tweets that matched the regex described in Section~\ref{sec:data_hidden_services}, resulting in 509,173 tweets. Next, we ran another pass on the matched tweets to group them into unique Bitcoin addresses. From 509,173 matched tweets, we found 4,183 unique addresses and identities, where an address of an identity appeared in 165 different tweets, on average. We refer to this list of addresses as the {\sffamily twitterUsers} dataset.

\paragraph{Forum} BitcoinTalk is one of the most popular Bitcoin forums with more than 900K users who exchange interests, technical expertise, and experiences in the development of the Bitcoin software. The forum also has several different sections for coin mining, technical support, and the economy of Bitcoin. It is the first forum of its kind that discusses topics related to Bitcoin and has reached its billionth post in Jul, 2012. As of Nov 2017, the forum contained around 2.5 billion posts. Based on its popularity, we sought to use it as a resource to extract public addresses of Bitcoin users.

In our experiment, we crawled and parsed 900K user profiles by retrieving each profile page using its URL, where each page is indexed by a unique user identifier that starts from 1. Overall, the crawling resulted in 22 GB of unparsed user profiles in HTML format. Having the profiles downloaded, we parsed them searching for Bitcoin addresses using the regex described in Section~\ref{sec:data_hidden_services}, resulting in 40,970 unique addresses and identities. We refer to this list of addresses as the {\sffamily forumUsers} dataset.

\subsection{Wallet-Closure Analysis}
\label{subsec:closure_analysis}
The goal of wallet-closure analysis is to expand the set of Bitcoin addresses that are controlled by a user in order to establish a unique many-one mapping between addresses and an identity. Increasing the number of Bitcoin addresses per user allows us to identify more links between the user and hidden services. Using the first wallet-closure heuristic originally proposed by Meiklejohn et al.~\cite{meiklejohn:2013}, we define the closure of a Bitcoin address as follows: If addresses $A$ and $B$ are in a closure, then there exists a transaction where addresses $A$ and $B$ appear as inputs. The motivation for this is that if two addresses appear in the same transaction as inputs, then they are likely to be controlled by the same user, since they are signed by the private keys of the owner who performed the transaction. However, this heuristic is noisy when users utilize mixing services or use CoinJoin transactions, as the mixing process results in closures that include addresses belonging to multiple identities. Accordingly, one might end up with closures that have a large number of addresses that are not mutually exclusive. In other words, there will be some Bitcoin address that appear in multiple closures.

Mixing services are third party services that receive coins from one user, mixes the coins with those received from other users, and then sends back the same amount of coins, albeit shuffled with different addresses using a number of transactions, to the original user. CoinJoin, on the other hand, is a peer-to-peer mixing protocol that achieves a similar goal, but uses a more sophisticated approach. These services are used to improve the anonymity of transactions and reduce linkability. In order to eliminate the effect of mixing services, we perform the following cleaning process: We find all closures that share at least one address and consequently merge them, after which we remove all closures that have been merged, ending up with unique closures that have no intersections and are mutually-exclusive. While doing so ensures that wallet-closures which belong to multiple users are not double counted, it also means that the resulting number of users that are linked to hidden services represents a lower-bound estimate of the actual number of user that can be linked and deanonymized.

\begin{table*}
{
\centering
\begin{tabular}{l r r r r r r l l r} \toprule
 & \multicolumn{3}{c}{\# linked users} & Volume & \multicolumn{2}{c}{Flow of money (\bitcoin)} & \multicolumn{3}{c}{Lifetime (dd/mm/yyyy)}\\
\cmidrule(lr){2-4}
\cmidrule(lr){6-7}
\cmidrule(lr){8-10}
Name & {\sffamily twitterUsers} & {\sffamily forumUsers} & Total & (\# txs) & Incoming & Outgoing & First tx & Last tx & \# days\\ \midrule
WikiLeaks & 11& 35 & 46 & 25,569 & 4,011.95 & 4,000.40 & 14/06/2011 & 24/07/2017 & 2,163\\
Silk Road & 4 & 18& 22 & 979 & 29,675.86 & 29,658.80  & 02/10/2013 & 24/07/2017 & 1,321\\
Internet Archives & 3 & 13 & 16 & 2,534 & 155.55 & 99.61 & 09/05/2013 & 24/07/2017 & 1,537\\
Snowden Defense Fund & 3 & 8 & 11 & 1,642 & 201.88 & 198.38 & 10/08/2013 & 15/07/2017  & 1,370\\
The Pirate Bay & 3 & 7 & 10 & 1,192 & 76.80 & 76.78  & 29/05/2013 & 08/07/2017 & 1,151\\
DarkWallet Mixer & 9 & 1 & 10 & 1,084 & 109.97 & 92.75 & 16/04/2014  & 02/11/2016 & 932\\
ProtonMail & 1 & 7 & 8 & 2,850 & 143.17 & 68.78  & 17/06/2014 & 24/07/2017 & 1,498\\
OpenStreetMap Donations & 0 &  5 & 5 & 440 & 24.01 & 24 & 13/05/2013 & 13/07/2017 & 1,522\\
Darknet Mixer & 1 & 2 & 3 & 22,110 & 306.16 & 341.48 & 21/01/2014 & 24/07/2017 & 1,645\\
Liberty Hackers & 0 & 2 & 2  & 85 & 2.79 & 2.79 & 10/04/2013 & 11/07/2017 & 1,553\\
Onion Mail & 1 & 0 & 1 & 226 & 84.92 & 84.92 & 15/09/2014 & 30/03/2015 & 196\\
Bitcoin Fog & 1 & 0 & 1 & 121 & 10.46 & 10.46 & 12/08/2014  & 09/01/2015 & 150\\
Bitmessage Mail & 0 & 1 & 1 & 106 & 2.78 & 1.37  & 28/04/2014 & 16/06/2017 & 1,145\\
Secure Tor Messaging & 0 & 1 & 1 & 105 & 12.5 & 12.5  & 02/01/2013 & 05/01/2016 & 1,098\\
Ransomware & 1 & 0 & 1 & 72 & 41.15  & 41.15 & 28/02/2014 & 15/08/2014 & 168\\
Bitcoin Lottery & 0 & 1 & 1 & 33 & 0.22 & 0.22  & 28/02/2014 & 02/05/2015 & 428\\
Libertarian Nuts  & 0  & 1 & 1 & 23 & 0.31  & 0  & 28/02/2014 & 24/10/2015 & 603\\
\midrule
Unknown1  & 1 & 1 & 2 & 42 & 8.32 & 8.31 & 23/06/2014 & 23/06/2017 & 1,096\\
Unknown2  & 1 & 0 & 1 & 132 & 4.19 & 4.18 & 22/06/2014 & 01/01/2015 & 193\\
Unknown3  & 0 & 1 & 1 & 39 & 7.95 & 7.94 & 01/05/2014 & 18/05/2017 & 1,113\\
\bottomrule
\end{tabular}
\caption{Linked hidden services. The three unknown services belong to onion pages that were taken down before we could manually identify and validate their hidden service provider.}
\label{table:hs_leaked}
}
\end{table*}

In our experiment, after applying wallet-closure, we were able to expand the {\sffamily twitterUsers} dataset by 619,006 additional addresses for 1,322 users out of the 4,183. The closures were more significant for the {\sffamily forumUsers} dataset, where we were able to add 18,508,012 addresses for 22,843 users out of the 40,970. In total, for the two datasets, we ended up with 19,172,171 addresses for 45,153 users, with an average of 425 addresses per user representing the average closure size. After wallet-closure cleaning, we ended up with 3,640 closures for the {\sffamily twitterUsers} dataset, and 23,567 closures for the {\sffamily forumUsers} dataset. These closures under-approximate user wallets, where each consists of at least one Bitcoin address and is uniquely mapped to its owner who is a user with an online identity. We use these closures to link users to hidden services in the next section.

Figure~\ref{fig:closure_cdf} shows the closure size CDFs for both datasets, before and after the cleaning process. As illustrated in the figure, there is a significant drop in the size of closures after cleaning; the median size decreased from 8 addresses to 4 for the {\sffamily twitterUsers} dataset, and from 103 addresses to 5 for the {\sffamily forumUsers} dataset. In fact, more than 90\% of the users in both datasets have 6 addresses or less in their wallets after cleaning. The figure also suggests that a larger number of BitcoinTalk users utilize mixing services than Twitter users, as shown by the difference in their before/after distributions.

\subsection{Bitcoin Transaction Analysis}
\label{subsec:api}
We now describe how we linked users to hidden services, give two deanonymization case studies, and analyze the economic activities of the linked services.

\subsubsection{Linking.}
In order to establish a link between a user wallet and a hidden service, we need to search the Blockchain for a transaction whose input is any of the addresses in the wallet and whose output is a hidden service address.

In our experiment, we first downloaded the whole Blockchain using the Bitcoin Core client software~\cite{bitcoin_core}, which is also responsible for managing the client's runtime and transactions. At the time of the analysis, the size of the Blockchain was over 230GB and it took nearly two days to download and sync the Blockchain on an average Internet connection. The Bitcoin Core client does not provide an easy, native way to access Blockchain transactions. For that, one can use APIs built on top of Bitcoin Core, such as Bitcore Node~\cite{bitcore_node}, which extends Bitcoin Core and provides additional indexing for more advanced address queries, and Insight API~\cite{insight_api}, which is a RESTful HTTP and web socket API service for Bitcore Node.

Using the Insight API on top of Bitcore Node, we performed the linking process after wallet-closure as follows: For each address in the {\sffamily hiddenServices} dataset, we queried the Blockchain for the transactions history of that address. This query returns a set of transactions in which the address appears as either an input or an output. We then issued the same query for each address in the {\sffamily twitterUsers} and {\sffamily forumUsers} datasets. This resulted in two sets of transactions; one for hidden services and one for users. Finally, we cross-matched the two sets of transactions; if any address of a user is found as an input in any transaction where a hidden service address appears as an output, then the user has a relationship with that hidden service, and thus, a link is established. For the {\sffamily twitterUsers} dataset, we were able to link unique 28 users to 14 hidden services using 167 transactions. Similarly, for the {\sffamily forumUsers} dataset, we were able to link unique 97 users to 20 hidden services using 115 transactions. Some of these users were linked to multiple hidden services, and a total of 20 unique hidden services were linked to users from the two datasets. As suggested by the results, although Twitter users were smaller in number compared to BitcoinTalk users, they were more active and had a larger number of transactions with hidden services. In fact, some of these users are ``returning customers,'' as they have performed multiple transactions with the same hidden services.

Table~\ref{table:hs_leaked} lists the hidden services sorted by how many users were linked to them. The list is topped by WikiLeaks, a service that publishes secret information provided by anonymous sources, with 46 linked users. This is followed by the Silk Road, a famous black market on the Dark Web, with transactions from 22 users whose input coins have been seized by the FBI. Although the wallet address of Silk Road was seized, it is still receiving transactions until recently. However, from our analysis, we observed that a number of transactions were performed prior to the seizure. Ranked fifth, The Pirate Bay, which is known for infringing IP and copyright laws by facilitating the distribution of protected digital content, was linked to 10 users. Other services listed in Table~\ref{table:hs_leaked} include hacking services, such as Liberty Hackers, mixing services, such as Darknet Mixer, and various secure mailing providers, such as ProtonMail.

\subsubsection{Deanonymization.}
As linked users have associated online identities, they could be deanonymized with different levels of certainty, depending on how much personally identifiable information they have shared on their social network's user profiles. We next focus on two case studies that illustrate this threat in more detail. It is important to note that we found a number of sensitive details about these users including location, gender, age and email addresses. However, due to ethical considerations, we only disclosed the information that demonstrate the privacy implications of this type of analysis.

The first case study is of a Pirate Bay hidden service user. The associated online identity indicates that the user is a middle aged man from Sweden. This user is particularly interesting because The Pirate Bay website was founded by a Swedish organization called Piratbyr{\aa}n. Furthermore, the original founders of the website were found guilty in the Swedish court for copyright infringement~\cite{wiki:The_Pirate_Bay}. Since then, the website has been changing its domain constantly, and eventually operated as a Tor hidden service. Therefore, the existing link between this user and The Pirate Bay through recent transactions could be incriminating.

In the second case, we focus on the Silk Road hidden service. As shown in Table~\ref{table:hs_leaked}, there are 22 users that had a link to Silk Road through transactions with seized Bitcoin addresses. These users are located across the world in countries such as India, Canada, and the USA. They include 4 males and 6 females of different ages that range between 13 and 42 years. The 18 users from the {\sffamily forumUsers} dataset were active on BitcoinTalk between 2013 and 2015, while 3 of them are still active in 2017. As for the 4 users from the {\sffamily twitterUsers} dataset, they posted an average of 45 tweets in 2014. One particularly interesting user is a young teenager from the USA. This user has been a registered BitcoinTalk member since 2013, and had a transaction with Silk Road in 2013, the takedown year, during which he was even younger than what his current age shows on his profile. The associated profile also includes his personal website, which contains links to his Facebook, Twitter, and Youtube profiles.

\subsubsection{Economic Activity}
\label{sec:economicActivity}
In order to gain insights about the economic activities of the linked hidden services, we analyzed all of their transactions in the Blockchain. In our experiment, for each service, we collected information about its number of transaction (i.e., volume), the amount of coins the service has received or sent (i.e., flow of money), and the timestamps of its first and last transactions (i.e., lifetime). The results are also summarized in Table~\ref{table:hs_leaked}. We refer the interested reader to Appendix~\ref{sec:appendix} for the economic activity analysis of the all hidden services, including the unlinked ones.

\paragraph{Volume.}
While the list of services is small, our results indicate that they have been involved in a relatively large number of transactions. For example, WikiLeaks tops the list with 25,569 transactions. The Darknet Mixer, on the other hand, has a volume of 22.1K transactions that is greater than the remaining services combined. One explanation for this popularity is that users are actually aware of the possibility of linking, and try to use mixing services in order to make traceability more difficult and improve their anonymity.

\paragraph{Money flow.}
We calculated the total incoming and outgoing Bitcoins for each hidden service in order to determine how much money is actually flowing in and out of their addresses. One interesting observation is that almost the same amount of coins flow in and out of these addresses. This indicates that the money is being distributed to other addresses, and is not stored on payment-receiving addresses. One explanation for this behavior is that by distributing funds to other addresses, a hidden service can reduce coin traceability. Also, hidden services still need to distribute their revenues among owners, sellers, and other stakeholders.

We also observed that multiple hidden services have a revenue of more than 4K Bitcoins and up to 29.6K, where one Bitcoin was valued at 19.7 thousand USD as of Dec, 2017. The Silk Road, for example, has received more than 580 million USD on its address.

\paragraph{Lifetime.}
Tracking the economic activity of hidden services over time allows us to estimate their operational lifetime, at least as seen from their associated addresses. From the transaction history of each hidden service, when filtered by its Bitcoin address, we define the lifetime of the service as the difference between the timestamps of the last and the first transactions involving the address. This allows us to estimate the period of operation of each hidden service, and accordingly, determine if the service is still active.

As summarized in Table~\ref{table:hs_leaked}, hidden services vary in their lifetime, ranging from 2--6 years of operation. We note that the first transaction date does not imply that the service began its operation on that date; it indicates the date on which the service started receiving Bitcoin payments. Looking at last transaction dates, most of the hidden services are still active in 2017.

\section{Discussion}
\label{sec:discussion}
We now discuss the deanonymization attack, focusing on its feasibility and implications. We also list the limitations of this work and highlight a number of existing countermeasures for improved privacy and anonymity.
\subsection{Feasibility}
\label{subsec:anonymity_attacks}
Tor is expected to maintain its anonymity guarantees even with an active, local adversary who controls a fraction of network resources, 20\% of the routers, for example. The goal of such an adversary is to link the source of a communication tunnel to its destination. In our setting, this maps to linking a hidden service user to a hidden service operator. By observing both ends of a circuit, even a passive adversary can confirm a connection exists between Alice and Bob using time and traffic analysis~\cite{tor}. If the adversary controls only a small fraction of the network, the chance of such an end-to-end compromise is very small. However, side-channel attacks can be used to increase the success rate of deanonymization attacks~\cite{arp2015torben}.

The results in Section~\ref{subsec:api} show that it is possible to deanonymize hidden service users without the need to control network resources nor inspect the traffic. As discussed in Section~\ref{sec:adversary_model}, we consider a passive, limited adversary that falls within the adversary model of Tor, and focus on hidden services that use Bitcoin as a payment method. Through a real-world experiment, we showed that an adversary can link users, who publicly share their Bitcoin addresses on online social networks, with hidden services, which publicly share their Bitcoin addresses on onion landing pages. This is accomplished by inspecting the transactions involving these two addresses in the Blockchain.

While collecting Bitcoin addresses of hidden services, we found that many services started to hide their payment Bitcoin addresses from onion landing pages. One explanation of this trend is that hidden service operators realized that publicly sharing Bitcoin addresses can leak information which could be used for linking and improving traceability. Instead, these services let users register accounts on their website and use them to perform transactions without exposing the addresses used to receive Bitcoins. The way this works is as follows: If Alice wants to perform a transaction with a hidden service, she starts by creating a personal account on the service. The hidden service creates and controls a new Bitcoin address, from a new public/private key pair, to which Alice makes a transaction from her personal Bitcoin addresses.

An active, more resourceful, local adversary can compile a larger set of hidden service addresses by performing small transactions with such services in order to reveal their addresses. Moreover, adversaries can impersonate a hidden service and receive payments on addresses they control. Deanonymization that exploit such techniques has been practiced in the wild by governments to uncover illegal hidden services~\cite{cochran_2013}. The adversary can also map collected Bitcoin addresses to IP addresses in order to deanonymize users at a more granular level~\cite{koshy_koshy_mcdaniel_2014}.

In this work, due to ethical concerns, we simulated only a passive, limited adversary. However, active, local adversaries that are well-funded are likely to exist in practice. We note that the success rate achieved by a passive, limited adversary represents a lower bound of that of an active, local adversary.

\subsection{Implications}
\label{sec:implications}
The main security implication of our work is that a Bitcoin addresses can be exploited to deanonymize users. The experiment we conducted can be extended to include other online social networks and information sources. It is also likely that an online identity, or a user account, on one social network has links to others networks that provide additional PIIs. Some users explicitly reveal their name, age, nationality, and other information in their bio or through posts. This represents a serious threat to their anonymity, since the hidden services they engage with might be associated with questionable transactions. The linking process can also be used as a tool by third-parties to track users, perform surveillance, and audit financial transactions.

Due to these security and privacy concerns, users have to follow simple yet effective guidelines in order to protect themselves. First, users should never expose their Bitcoin addresses along with their personally identifiable information. Second, as discussed in Satoshi's white paper~\cite{nakamoto:2008}, a new address should be generated for each transaction in order to reduce linkability of transactions, regardless of whether the user is the sender or the receiver of the payment. This is especially true for cases where users expose a donation address on different kinds of online social networks. Unfortunately, a large number of users do not follow this practice, possibly due to poor usability of Bitcoin tools, unfamiliarity of Bitcoin internals, or reliance on third-party wallets and exchange services~\cite{krombholz2016other}.

\subsection{Limitations}
Our work has two main limitations. First, in our analysis, we assume that linking a user, represented as an online identity, to a hidden service is sufficient to deanonymize the user. However, this is not always true. Users can always create fake online identities in order to hide their real ones. While doing so improves their anonymity, once the links are established the adversary can perform online surveillance to track down the users and uncover their true identities. The second limitation is related to the use of mixing services. While the wallet-closure cleaning process we used eliminates the effect of mixing, it is aggressive and can exclude users who did not use mixing services at all. Accordingly, our results under estimates the prevalence of the deanonymization threat.

It is important to mention that some of the linked users through this analysis are not concerned about their transactions anonymity to services such as Internet Archives. However, some users are unconscious about protecting their privacy, and this type of analysis can be used to incriminate these users. For example, services such as The Pirate Bay are illegal in Sweden, and therefore, establishing a link through a transactions with such a service could potentially be used as an evidence to accuse the user of illegal activities.

One might argue that the number of deanonymized users is small. However, in order to understand the significance of the results, it is useful to put the numbers into perspective. According to recent web statistics~\cite{internet_users,tor_metric}, the number of worldwide Internet users is around 3.58 billion and the number of Tor network users is about 3 million, which are a superset of Tor hidden service users. This means that 0.086\% of Internet users are Tor network users, on average. The datasets we collected include 45,153 Twitter and BitcoinTalk users, out of which 0.277\% of them were deanonymized. While this percentage is larger than 0.086\%, likely due to biased sampling, it is still relatively small, as expected. In other words, because we do not know how many of the users in the datasets are Tor hidden service users, we should expect that only a small percentage of them can be deanonymized.

\subsection{Countermeasures}
\label{sec:countermeasures}
There are two general ways to achieve improved anonymity for users and hidden services. The first one is operational, and it focuses on following Bitcoin best practices, as discussed in Section~\ref{sec:implications}. For those users who can be linked, the best course of action for them is to clean their social network footprint, focusing on removing PII that is publicly shared or deleting their linked online identities, all together. The second way to improve anonymity is technical, and it involves improvements to the current Bitcoin protocol or the introduction of new crypto-currencies that are based on Bitcoin's Blockchain technology.

Second generation anonymization techniques, such as CoinJoin, Fair Exchange~\cite{jayasinghe2014optimistic}, CoinSwap, and stealth addresses have been proposed to be implemented as extensions or services for Bitcoin's original protocol. These are discussed in details in~\cite{moseranonymous}. Furthermore, other alternative coins based on different modifications to Bitcoin protocol have been introduced to provide additional anonymity for transactions on the Blockchain. The most prominent of them being Monero, which is based on Crypto Note v2.0 protocol~\cite{saberhagen2013crypto} and Zcash~\cite{zerocash}. In fact, Monero is already making its way into the hidden services of the Dark Web~\cite{hs_monero}. These new coins aim to deliver full anonymity of transactions, and have their own advantages and disadvantages compared to Bitcoin.

\section{Related Work}
\label{sec:related_work}
There are a number of studies that investigate user anonymity and privacy concerns in Bitcoin~\cite{Reid:2013,meiklejohn:2013,DuPont:2015,fleder:2015}. Fergal and Martin~\cite{Reid:2013} demonstrated that using passive analysis of publicly-available Bitcoin information can lead to a serious information leakage. They constructed two networks representing transactions and users from the Blockchain. Integrating these networks with off-network information, such as user profiles from online social networks, and context discovery and flow analysis techniques, it was possible to study the flow of Bitcoins between addresses and investigate thefts. Fleder et al.~\cite{fleder:2015}, on the other hand, explored the level of anonymity in the Bitcoin network. The authors annotated the transaction graph by linking user pseudonyms to online identities collected from online social networks. They also developed a graph-analysis framework to summarize and cluster the activity of users. The analysis links identities of users to their transactions. These studies form the base for our approach, as we use some of their techniques in our analysis. The difference in our study is that we target a specific portion of Bitcoin users, which are Tor hidden service users. We also study the economic activities of linked hidden services, which is important to understand the level of threat the users are exposed to.

DuPont and Squicciarini~\cite{DuPont:2015} proposed a technique to determine a Bitcoin user's physical location by examining user spending habits and linking it to the user's time zone. Androulaki et al.~\cite{Androulaki2013} studied the privacy provisions in Bitcoin through a simulation mimicking the use of Bitcoin as the digital currency for daily transactions in a typical university setting. The study shows that behavior-based clustering can unveil the profiles of 40\% of Bitcoin users even if they are using recommended privacy measures. This method can be used with our techniques to increase the deanonymization level from an online identity to a physical identity.

A recent study by Harrigan and Fretter~\cite{harrigan2016unreasonable} showed the effectiveness of address clustering using the Blockchain of Bitcoin. These clusters are constructed using different heuristics such as the one we used in our study. The authors performed address clustering on the entire Bitcoin's Blockchain and showed that despite the existence of CoinJoin and mixed transactions, address clustering is still suitable for Blockchain analysis and re-identification attacks. The findings presented in their work strengthen our analysis results and further proves that our linking to hidden services is still valid until the current time.

\section{Conclusion}
\label{sec:conclusion}
We show that using Bitcoin as a payment method for Tor hidden services leaks information that can be used to deanonymize their users. This represents a serious threat to these users, because they actively seek to maintain their anonymity by using Tor. The deanonymization is mainly due to the lack of retroactive operational security present in Bitcoin's pseudonymity model. In particular, by inspecting historical transactions in the Blockchain, an adversary can link users, who publicly share their Bitcoin addresses on online social networks, with hidden services, which publicly share their Bitcoin addresses on their onion landing pages.

In a real-world experiment, we were able to link many users of Twitter and the BitcoinTalk forum to various hidden services, including WikiLeaks, Silk Road, and The Pirate Bay. Using information from their public user profiles, we were able to show concrete case studies where the anonymity of the users is broken. Our results has one immediate implication: Bitcoin addresses should always be assumed compromised as they can be used to deanonymize users.

\section*{Acknowledgement}

We would like to thank Hasan Al Jawaheri, Ryan Riley, Jaideep Srivastava, and the people at the Cybersecurity Initiative for Blockchain Research (CIBR) for their help and feedback.\footnote{For latest research outcomes, please visit \url{https://qcri.github.io/cibr}} The first author would like to thank Qatar University for a generous research assistantship. This work was partially supported by the NPRP grant 7-1469-1-273 from Qatar National Research Fund. The statements made herein are solely the responsibility of the authors.

\bibliographystyle{refs}
\bibliography{refs}

\appendix
\section{Appendix}
\label{sec:appendix}

In what follows, we present a high-level analysis of the economic activity involving  all hidden services from the {\sffamily hiddenServices} dataset.

\subsection{Volume}

\begin{figure}[H]
\centering
\includegraphics[width=0.65\linewidth]{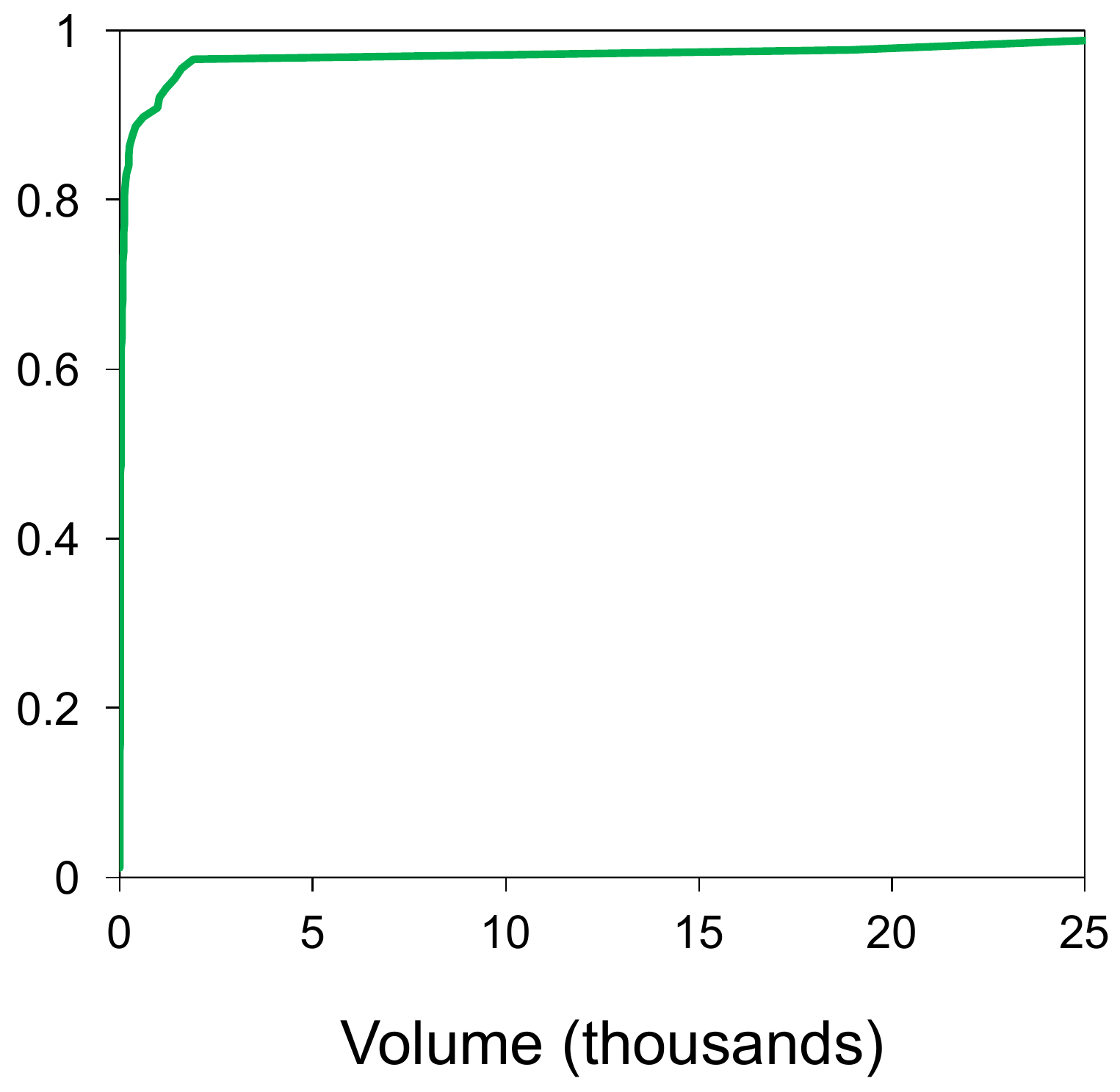}
\caption{Number of transactions of hidden services}
\label{fig:hsVolume}
\end{figure}

Figure~\ref{fig:hsVolume} shows the CDF of the total volume from all services, as defined in Section~\ref{sec:economicActivity}. We can see from the figure that 10\% or less hidden services had more than 1,000 transactions. This complies with our previous results showing that most of the transactions from {\sffamily twitterUsers} and {\sffamily forumUsers} datasets were attributed to the top 4--5 hidden services, in terms of volume.

\subsection{Money Flow}

\begin{figure}[H]
\centering
\includegraphics[width=0.65\linewidth]{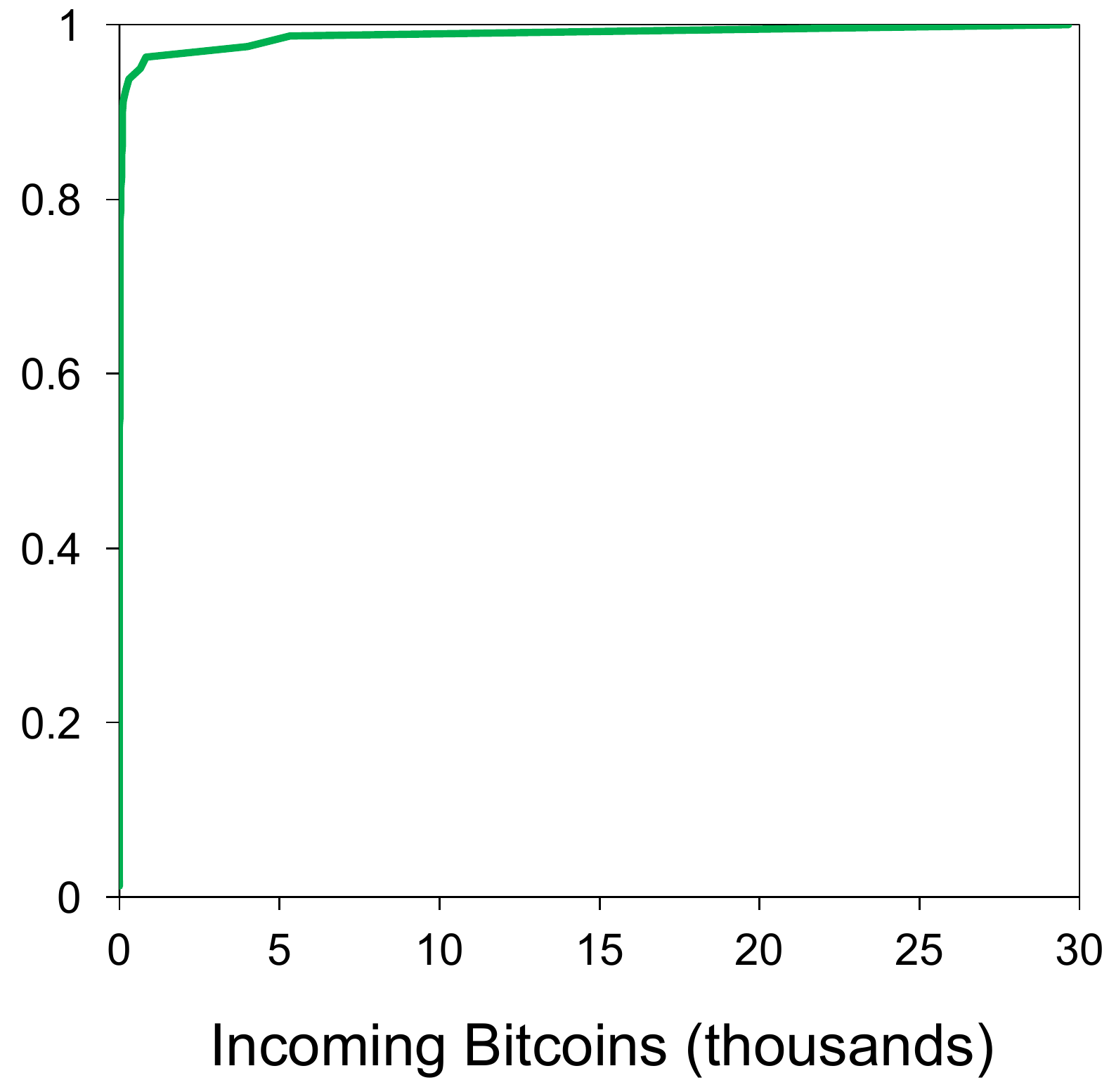}
\caption{Money flow for hidden services}
\label{fig:hsIncome}
\end{figure}

The money flowing in and out of hidden service addresses is almost identical, so is the corresponding money flow CDFs. Figure~\ref{fig:hsIncome} shows the CDF of the total incoming Bitcoins received by all services. As the figure shows, only the top 10\% of hidden services had received more than 100\bitcoin. The distribution is relatively skewed due to the fact that the top-3 services had significantly bigger revenue, with more than 4,000\bitcoin.

\subsection{Lifetime}

\begin{figure}[H]
\centering
\includegraphics[width=0.75\linewidth]{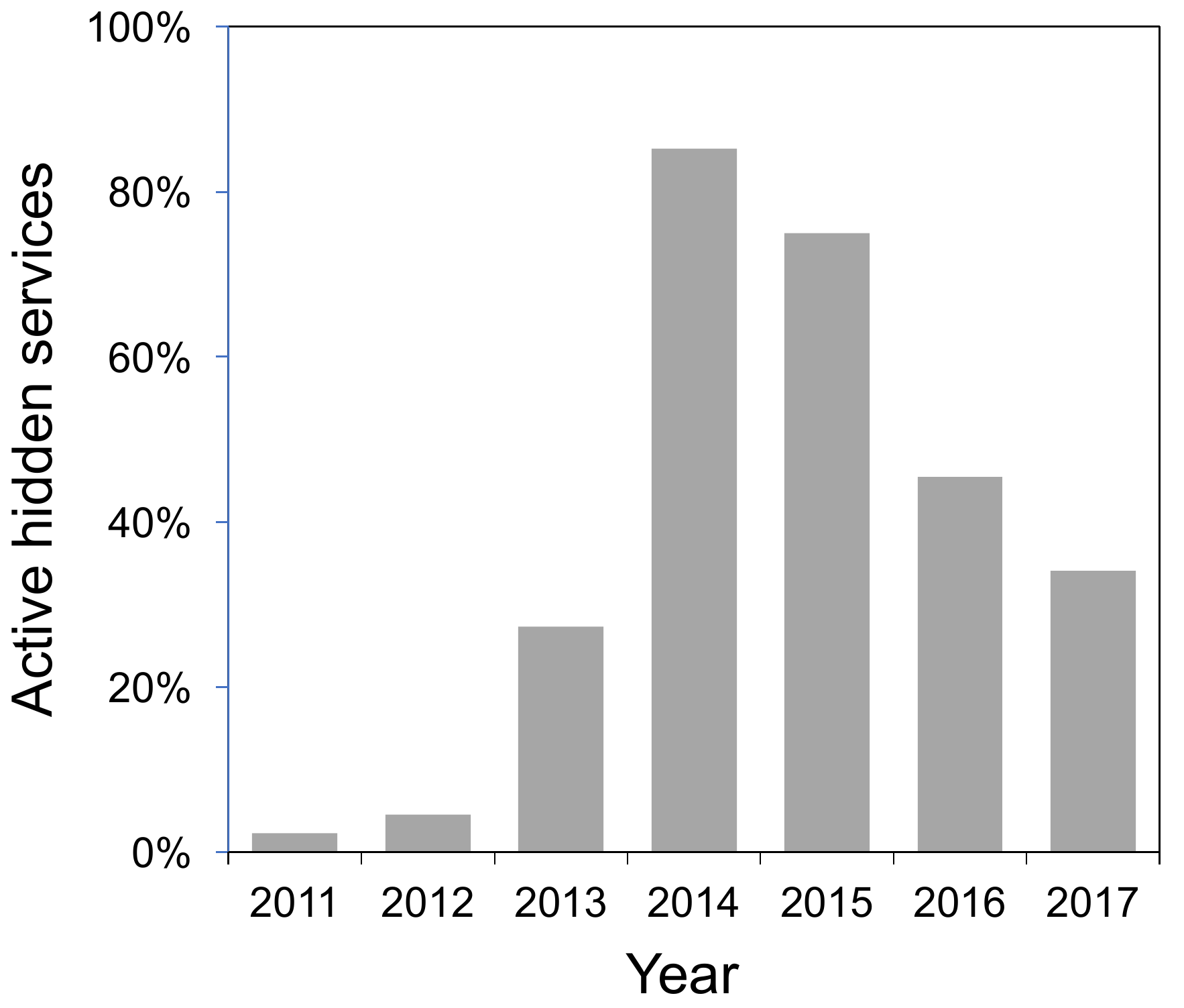}
\caption{Lifetime of hidden services}
\label{fig:hsLifetime}
\end{figure}

Figure~\ref{fig:hsLifetime} shows the percentage of hidden services that were active during the period 2011--2017. As denoted by the figure, most of these services  were active during the years 2014--2015, which is the same time frame we based our data collection period on. Furthermore, as expected, the activity for their addresses is significantly lower in the following years. This is most likely due to the usage of online wallets or their migration to other cryptocurrencies, as discussed in Section~\ref{sec:discussion}.

\end{document}